\documentstyle[sprocl,epsf]{article}

\begin{document}

\title{Minkowski Functionals in Cosmology: An Overview}

\author{Jens Schmalzing}

\address{Ludwig--Maximilians--Universit\"at, Theresienstra{\ss}e~37,
80333~M\"unchen, Germany \\
Max--Planck--Institut~f\"ur~Astrophysik,
Karl--Schwarzschild--Stra{\ss}e~1, 85740~Garching, Germany}

\maketitle

\abstracts{ Minkowski functionals  have recently  been introduced into
cosmology as novel tools for studying the large--scale distribution of
matter in the  Universe.  We present a brief  overview of  the method,
including its mathematical  foundations as well  as some completed and
upcoming applications.  }

\section{Mathematical background}

Useful  insights into the large--scale  structure  of the Universe are
possible by  investigating  the morphological   properties of  density
fields         and        point               sets.           Integral
geometry~\cite{schneider:brunn,weil:stereology}      provides       the
mathematical    framework   for  such   considerations.     Hadwiger's
Theorem~\cite{hadwiger:vorlesung} states  that in order  to measure the
morphology  of    a  $d$--dimensional  pattern,    the $d+1$ Minkowski
functionals are sufficient.

\section{Application to point processes}

Originally~\cite{mecke:robust}, applications of  Minkowski  functionals
in cosmology focussed on the Boolean grain model of a point set given,
for example,  by the coordinates of  galaxies  in redshift space.  The
Boolean model  constructs the union  set of balls centered  around the
members of  the point set.  Their common  radius can be  employed as a
diagnostic parameter, as the Minkowski   functionals of the union  set
change in  a characteristic fashion  with increasing radius.  

\begin{figure}
\begin{minipage}{5.5cm}
\epsfxsize=5.5cm\epsffile{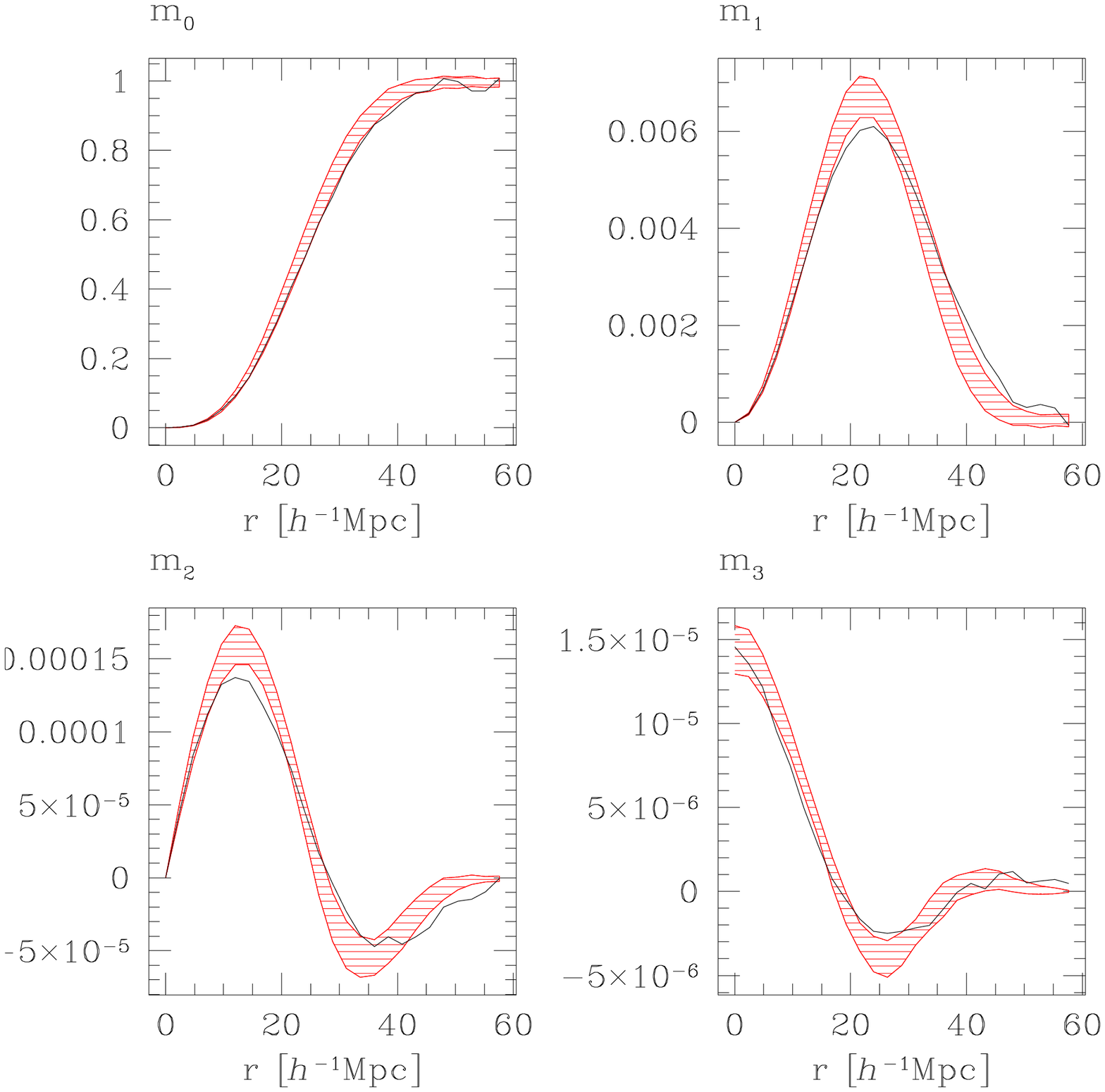}
\caption{
\label{fig:abellaco}
This plot shows  the Minkowski functionals  for the Abell/ACO  cluster
catalogue (solid line) compared  to several realizations of SCDM  mock
catalogues (shaded area).  }
\end{minipage}
\hfill
\begin{minipage}{5.5cm}
\epsfxsize=5.5cm\epsffile{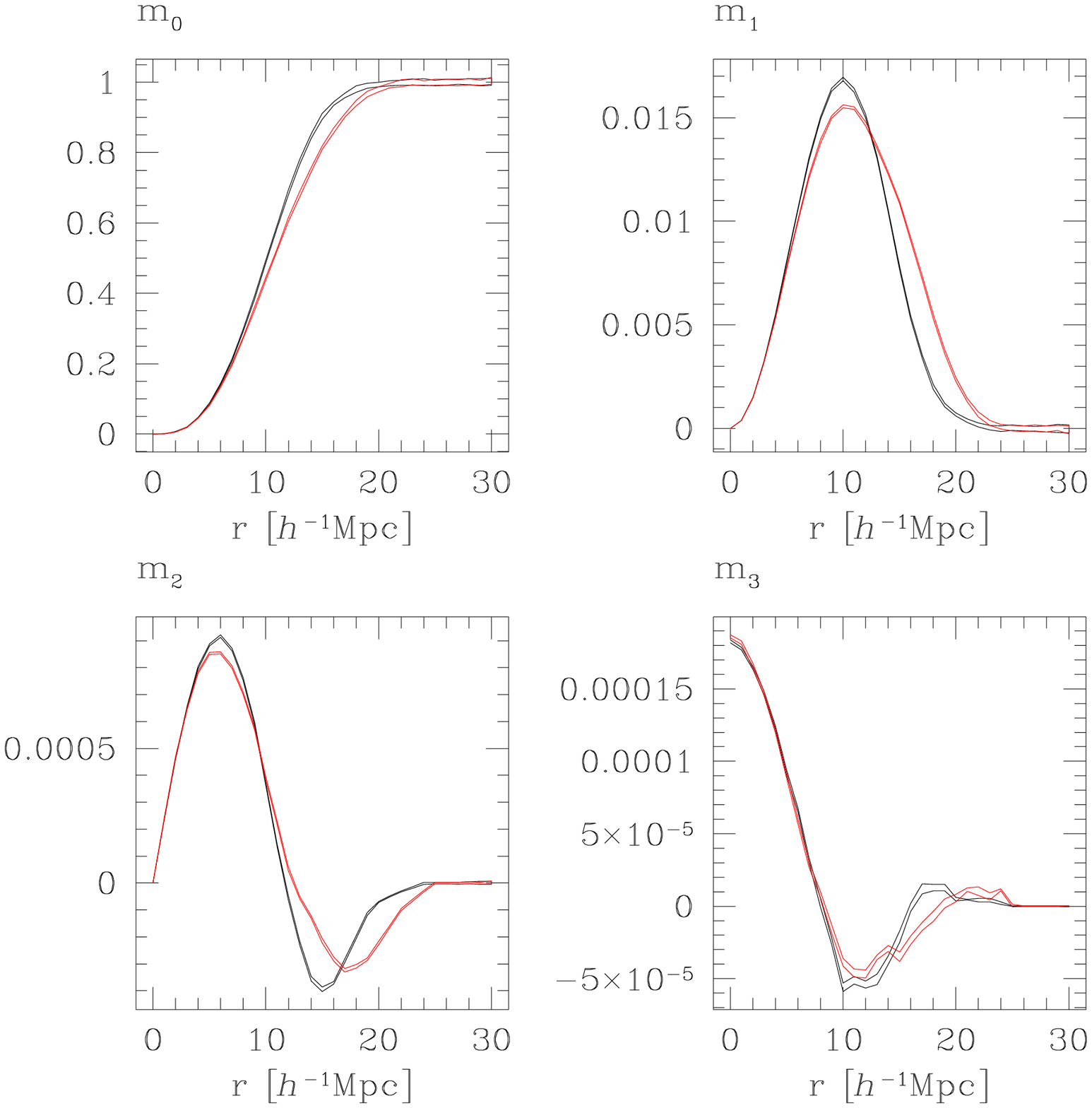}
\caption{
\label{fig:iras}
Here    we   display the  Minkowski    functionals  of volume--limited
subsamples  of  100$h^{-1}$Mpc depth  from the   northern and southern
hemispheres of the IRAS~1.2Jy galaxy catalogue. }
\end{minipage}
\end{figure}

The comparison of  a subsample of Abell and  ACO  clusters and several
varieties of   the  Cold  Dark Matter   scenario~\cite{kerscher:abell}
demonstrates   the   discriminative  power of  Minkowski  functionals.
Figure~\ref{fig:abellaco} clearly  shows the  failure of the  Standard
Cold Dark  Matter model.  Other  Cold  Dark Matter scenarios,  such as
Broken Scale Invariance or an  additional cosmological constant,  lead
to much better agreement.

Significant fluctuations  on large scales  are found in an analysis of
the IRAS~1.2Jy galaxy  catalogue~\cite{kerscher:fluctuations}.   Since
the  results  shown    in   Figure~\ref{fig:iras}   for   samples   of
100$h^{-1}$Mpc depth remain   robust on even  larger scales  and under
various tests, they are probably a  feature of the large--scale matter
distribution in the Universe, rather than  an effect in the catalogue.
An  explanation,  for example in  terms of  cosmic variance,  is still
lacking, and a topic of ongoing study.

\section{Outlook on further applications}

One of the Minkowski functionals is the Euler characteristic or genus,
which   has long been  used   in cosmology~\cite{gott:sponge}.  As  an
extension, all Minkowski functionals of a smoothed random field may be
considered~\cite{schmalzing:beyond}.   Furthermore, applications    to
cosmic          microwave              background           anisotropy
maps~\cite{schmalzing:minkowski_cmb} are possible.


\section*{References}

\small
\begin{thebibliography}{10}

\bibitem{schneider:brunn}
R.~Schneider.
\newblock {\em Convex bodies: the {B}runn--{M}inkowski theory}.
\newblock Cambridge 1993.

\bibitem{weil:stereology}
W.~Weil.
\newblock In {\em Convexity and its applications}, pp.~360--412. Basel 1983.

\bibitem{hadwiger:vorlesung}
H.~Hadwiger.
\newblock {\em Vorlesungen {\"u}ber {I}nhalt, {O}berfl{\"a}che und
  {I}soperimetrie}.
\newblock Berlin 1957.

\bibitem{mecke:robust}
K.~R. Mecke, T.~Buchert \& H.~Wagner.
\newblock {\em A\&A} {\bf 288} (1994) 697--704.

\bibitem{kerscher:abell}
M.~Kerscher et~al.
\newblock {\em MNRAS} {\bf 284} (1997) 73--84.

\bibitem{kerscher:fluctuations}
M.~Kerscher, J.~Schmalzing, T.~Buchert \& H.~Wagner.
\newblock {\em A\&A} submitted, astro-ph/9704028, 1997.

\bibitem{gott:sponge}
J.~R.~{Gott III}, A.~L. Melott \& M.~Dickinson.
\newblock {\em ApJ} {\bf 306} (1986) 341--357.

\bibitem{schmalzing:beyond}
J.~Schmalzing \& T.~Buchert.
\newblock {\em ApJ} {\bf 482} (1997) L1--L4.

\bibitem{schmalzing:minkowski_cmb}
J.~Schmalzing \& K.~M. G{\'o}rski.
\newblock {\em MNRAS} submitted, astro-ph/9710185, 1997.

\end{thebibliography}
\end{document}